# WAVE UNIVERSE
# AND SPECTRUM OF QUASARS REDSHIFTS

**A.M. Chechelnitsky,** Laboratory of Theoretical Physics,
Joint Institute for Nuclear Research,
141980 Dubna, Moscow Region, Russia
E'mail: ach@thsun1.jinr.ru

## ABSTRACT

In the framework of the Wave Universe concept it is shown, that the genesis of redshifts can be connected with the intra-system (endogenous) processes, which take place in astronomical systems. The existance of extremal redshift objects (quasars - QSO) with most probable

$$z = 3.513\ (3.847);\ \ 4.677;\ \ 6.947\ (7.4);\ \ 10.524;\ \ 14.7;\ \ 27.79;$$

is predicted.

## THE WAVE (MEGAWAVE) ASTRODYNAMICS CONCEPT

A wide set of yet noninterpreted (enigmatic from the point of view of standard paradigma of celestial mechanics and astrophysics [1,2]) observed and experimental data, connected with the dynamical structure and geometry of the Solar system (in particular, with the arragement of planetary, satellite orbits, distribution its velocities, etc.) and other astronomical systems can be adequately interprated in the framework of Wave (Megawave) Astrodynamics (and Wave Universe concept) [2-6].

Accordingly to these representations, real objects observed in the Universe (in the megaworld, such as astronomical systems, for example, the Solar system) appear principally *wave* dynamic systems (WDS), in the some sence similar to the atom system (Micro - Mega analogy [2]), and can be described by Fundamental wave equations (in particular, Schrodinger-type equation) (Fig.1).

The unique dimensional parameter $đ$, which enters into such a wave equation, has the dimension of sectorial velocity (circulation) [$cm^2/s$] and corresponds to characteristic scale of system (Co-dimensional principle [2]).

For the atom it has the order

$$đ = đ_e = \hbar/m_e = 1.15767\ cm^2 \cdot s^{-1}$$

($\hbar = 1.054572 \cdot 10^{-27}\ g \cdot cm^2 \cdot s^{-1}$ - Planck's constant, $m_e = 9.109389 \cdot 10^{-27}\ g$ – mass of electron), for the Solar system (SS)

$$đ = đ_{SS} \approx 10^{19}\ cm^2 \cdot s^{-1}.$$

## EXTREMELY LOW MASS

From the comparison of the circulation parameters, carried out in the end of 70s in the monograph [2, p.245], naturally follows an evident, lying at surface, consequence.

Representing the Solar system constant

$$đ = đ_{SS} = \hbar_{SS}/m_{SS}$$

like as for atom ($đ = đ_e = \hbar/m_e$), it is easy, for example, in the case $\hbar_{SS} = \hbar$, to obtain the representation

$$đ = đ_{SS} = \hbar_{SS}/m_{SS}$$

and the order of mass

$$m_{SS} = \hbar/đ_{SS} \approx 1.054572 \cdot 10^{-27}/10^{19} \approx 10^{-46}\ g.$$

The physical sence of appearance of such an extremally low mass merits the special discussion. Meanwhile just note that this valuation is close to the upper limit of the experimental valuation of the photon mass [7]:

$$m_\gamma < 9 \cdot 10^{-10}\ eV/c^2 = 1.604 \cdot 10^{-42}\ g\ (Ryan, 1985),$$
$$4.73 \cdot 10^{-12}\ ev/c^2 = 0.8432 \cdot 10^{-44}\ g\ (Chernikov, 1992),$$
$$1.0 \cdot 10^{-14}\ ev/c^2 = 1.782 \cdot 10^{-47}\ g\ (Williams, 1971),\ \ etc.$$

## SPECTRUM OF ELITE VELOCITIES

The Fundamental wave equation, described the Solar system (similarly to the atom system), separates the spectrum of physically distinguished, stationary - *elite* - orbits, corresponding to mean quantum numbers N, including the spectrum of permissible *elite* velocities $v_N$.

The following representation holds for the physically distinguished – *elite* – velocities $v_N$ in the

$G^{[s]}$ Shells of wave dynamical (in particular, astronomical) systems [3-6]:

$$v_N = v_N^{[s]} = (2\pi)^{1/2} \cdot C_*^{[s]}/N, \quad C_*^{[s]} = C_*^{[1]} \cdot \chi^{-(s-1)}, \quad s = 0, \pm 1, \pm 2, \ldots$$

($\chi$ - Fundamental parameter of Hierarchy (Chechelnitsky Number) $\chi = 3.66(6)$,
$C_*^{[1]}$ – Sound velocity of cosmic plasma in $G^{[1]}$ Shell [Chechelnitsky, [2-6], 1980-1992]),
where elite values $N$ (as it follows from observations) are close, in the general case, to the counting set of $N$ - *Integer (semi - integer).*

The most stable - *dominant* (strong elite) – orbits and the related *dominant* velocities correspond to the *dominant* values of quantum numbers, close to

$$N = N_{Dom} = 8; 11; 13; (15.5)\ 16; 19.5; (21.5)\ 22.5.$$

It can be shown, that

$$N_{TR} = \chi (2\pi)^{1/2} \cong 9.191$$

also is the physically distinguished (dominant) value N [6].

## INVARIANCE (UNIVERSALITY) OF THE ELITE VELOCITIES SPECTRUM

The spectrum of physically distinguished elite velocities $v_N$ and quantum numbers $N$ of arbitrary wave dynamic systems (WDS) has some universal peculiarity. It is practically identical - *invariant (universal)* for all known observed systems of the Universe.

In particular, the velocity spectra of experimentally well investigated Solar and satellite systems practically coincide for the observed planetary and satellite - dominant - orbits, corresponding to some (dominant) values of quantum numbers $N_{Dom}$. Thus it can be expected that the spectrum of elite (dominant - planetary) velocities of the Solar system (well identified by observations) can be effectively used as quite representative - *internal (endogenous)* - spectrum of elite (dominant) velocities, for example, *of far* astronomical systems of the Universe.

## PHYSICALLY DISTINGUISHED REDSHIFTS

In the framework of the developed representations of Wave (Megawave) Astrodynamics and Wave Universe concept [2-6] the analytical representation can be obtained for physically distinguished - preferably observed (elite) - redshifts of far astronomical objects (galaxies, quasars).

The physically justified by experience (and correct consequences) relation $z = f(v)$ between the velocity $v$ and the redshift $z$ has the form

$$z = \beta^2 = (v/c)^2, \qquad \beta = v/c,$$

where $c = 299792.458$ km·s$^{-1}$ - light velocity.

This correlation between the redshift $z$ and the (orbital) velocity $v$ (as opposed to other relations) is carefully examined experimentally in laboratory conditions - on the Earth (Paund and Rebka's experiment) and in Space - from the Sun (Brault's experiment) [8].

It is also interesting to note that the used square dependence in the functional (mathematical) plane is in fact identical to the relation used in the calculation of the so-called gravity redshift [8]

$$z = GM/c^2 r = (v/c)^2,$$

where $v^2 = GM/r$, $v$ - orbital velocity.

## THE PEAKS z IN OBSERVATIONS AND IN THEORY

It may be shown that the most important peaks in histograms (of distribution) of the observed $z$ unaccidentally and with sufficient reliability coincide with the physically distinguished-dominant - values $z_N^{[s]}$ (at $N = N_{Dom}$).

For example, the peaks widely known from observations peaks (Figs. 2, 3) from [9-10]

$$z \approx 2,\ z \approx 1,\ z \approx 0.5,\ z \approx 0.35\ \text{(and other)}$$

coincide with the dominant ($N = N_{Dom}$) $z_N^{[-6]}$ values of $G^{[-6]}$ Shell

$$z_N^{[-6]} = z_*^{[-6]} \cdot 2\pi/N^2, \qquad z_*^{[-6]} = (C_*^{[-6]}/c)^2 = [(C_*^{[1]}/c) \cdot \chi^7]^2,$$

$$z_N^{[-6]} = 2.067;\ 1.093;\ 0.782;\ (0.55)0.516;\ 0.347;\ (0.286)0.261$$

and also (for $N_{TR} = 9.191$) $z_{TR}^{[-6]} = 1.57$.

## ENDOGENOUS NATURE OF z

The set of large quantity of facts, agreement between of theory and observations, including the possibility of correct description of distinguishing peaks $z$ over all the observed redshifts range (beginning from $z=0$) makes the next conclusion natural.

**Assertion.** It seems very probable that the true genesis and physical nature of the observed redshifts is considerably closer connected with *the own (inner)* wave shell structure of astronomical systems (galaxies, quasars), than with the "kinematic" motion (translation) of their mass center - with the galaxies "expansion".

## ABOUT THE EXISTENCE OF OBJECTS WITH EXTREMAL z

In the framework of the Wave Universe representations it must be expected that replenishing statistics of newly discovered astronomical objects will be characterized by the distribution peaks at z that correspond to the physically distinguished - *dominant* - values of redshifts, in particular, belonging to the $G^{[-7]}$ Shell:

$$z_N^{[-7]} = z_*^{[-7]} \cdot 2\pi/N^2, \quad z_*^{[-7]} = (C_*^{[-7]}/c)^2 = [(C_*^{[1]}/c) \cdot \chi^8]^2 = 283.08668,$$
$$z_N^{[-7]} = 27.79; 14.7; 10.524; (7.4)6.947; 4.677; (3.847)3.513.$$

Already at the present it is interesting to note apparently, unaccidental compliance of the observed values z of (remotest for 1986) quasars z=3.53 (quasar OQ 172) and z=3.78 (quasar PKS 2000-330) with the pointed above $z^{[-7]}$ dominant values of the $G^{[-7]}$ Shell (z=3,513 and z=3.847).

Thus, it is not excluded, that the quasars OQ 172 and PKS 2000 - 330 will become not as much *the last* from discovered quasars of preceding population QSO ($G^{[-6]}$) with active $G^{[-6]}$ Shell as *the first* (and evidently having not the highest values of z) from discovered quasars of *new population* QSO ($G^{[-7]}$) with active $G^{[-7]}$ Shell.

## THE PROBLEM OF SEARCH

Basing on the above - discussed prognosis, we may also point to a set of supplementary physical orientating circumstances, that essentially shorten the search field for the objects with extermal z. One of them resides in the fact that the search must be carried out, in particular, among the astronomical objects having abundant radiation (peculiarities, peaks, radiation anomalies), besides gamma, in close infrared range, too. Really, for example, for hydrogen $L_\alpha$ - line

$$\lambda(L_\alpha) = 1215.67 \text{Å} = 121.567 \text{ nm} = 0.121567 \text{ } \mu m$$

we have the system of shifted (by the redshift $z=z^{[-7]}$) wave lenghts

$$\lambda_N^{[-7]} = \lambda(L_\alpha)(1+z_N^{[-7]}) = 3.50; 1.90; 1.40; (1.02) 0.966; 0.69; (0.589) 0.548 \text{ } \mu m$$

that lay in *IR*-range.

Purposeful search of objects (most probably having z that are close to the pointed above), in particular, among objects as Seyfert galaxies, Markarjan galaxies, may lead to discovery of new astronomical systems, which are characterized by extremal, so far unknown values of redshifts.

## FOLLOWING OBSERVATIONS

Three years after the exposition of preceding results in 1986 [11-12] followed by a discussion between a confined circle of researchers - astrophysicists, in the end of 1989, american scientists from the Palomar Observatory M. Schmidt, J. Gunn, D. Schnaider discovered the extremely far object of the Universe - quasar in the Ursa Major constellation. It is interesting to note also that using of the experimental "solar" value N=19.43 (instead of N=19.5) indicates the more close (to discovered) value z=4.71 (instead of z=4 .677).

---

**Post Scriptum (2000)** [From Chechelnitsky, 2000]:

**What Quasars with Record Redshifts Will be Discovered in Future? Megaquantization in the Universe.**

It is clear, *Megaquantization* (quantization "in the Large"), observed megaquantum effects are not monopolic privelege of only Solar system.

Let us point the brief resume of research (prognosis), connected with problem of redshift quantization of far objects of Universe – quasars (QSO) [Chechelnitsky, (1986) 1977]:

"*Abstract*: In the framework of the Wave Universe concept it is shown that the genesis of redshifts can be connected with the *intra-system (endogenous)* processes which take place in astronomical systems. The existence of extremal redshift objects (quasars – QSO) with most probable z=3.513 (3.847); 4.677; 6.947 (7.4); 10.524; 14.7; 27.79; … is predicted."

Prognosis already had justified successively for extremal values of z redshifts

$z_{theory}$ = 3.513,   $z_{obs}$ = 3.53 (quasar OQ172)
$z_{theory}$ = (3.847),  $z_{obs}$ = 3.78 (quasar PKS2000-330)
$z_{theory}$ = 4.677,   $z_{obs}$ = 4.71 (Schmidt, Gunn, Schnaider, 1989)
            $z_{obs}$ = 4.694 (4.672) (quasar BR1202-0725, Wampler et al., 1996)

At the present time, apparently, also the object Q2203+29 G73 with record value z of redshift z=6.97 is discovered in special Astrophysical Observatory (SAO, Russia) $z_{theory}$ = 6.947, $z_{obs}$ = 6.97 (Q2203+29 G73, Dodonov et al., 2000).

The Quene – for objects with even more high redshifts z = 10.524; 14.7; …

Consequences of such successfully realizable prognosis, imperatives of observations not only are unexpected for the Standard cosmology, but also, probably, its can stimulated the radical reconsideration of many habitual representations, having become as freezen dogmas.

**Figure 1**

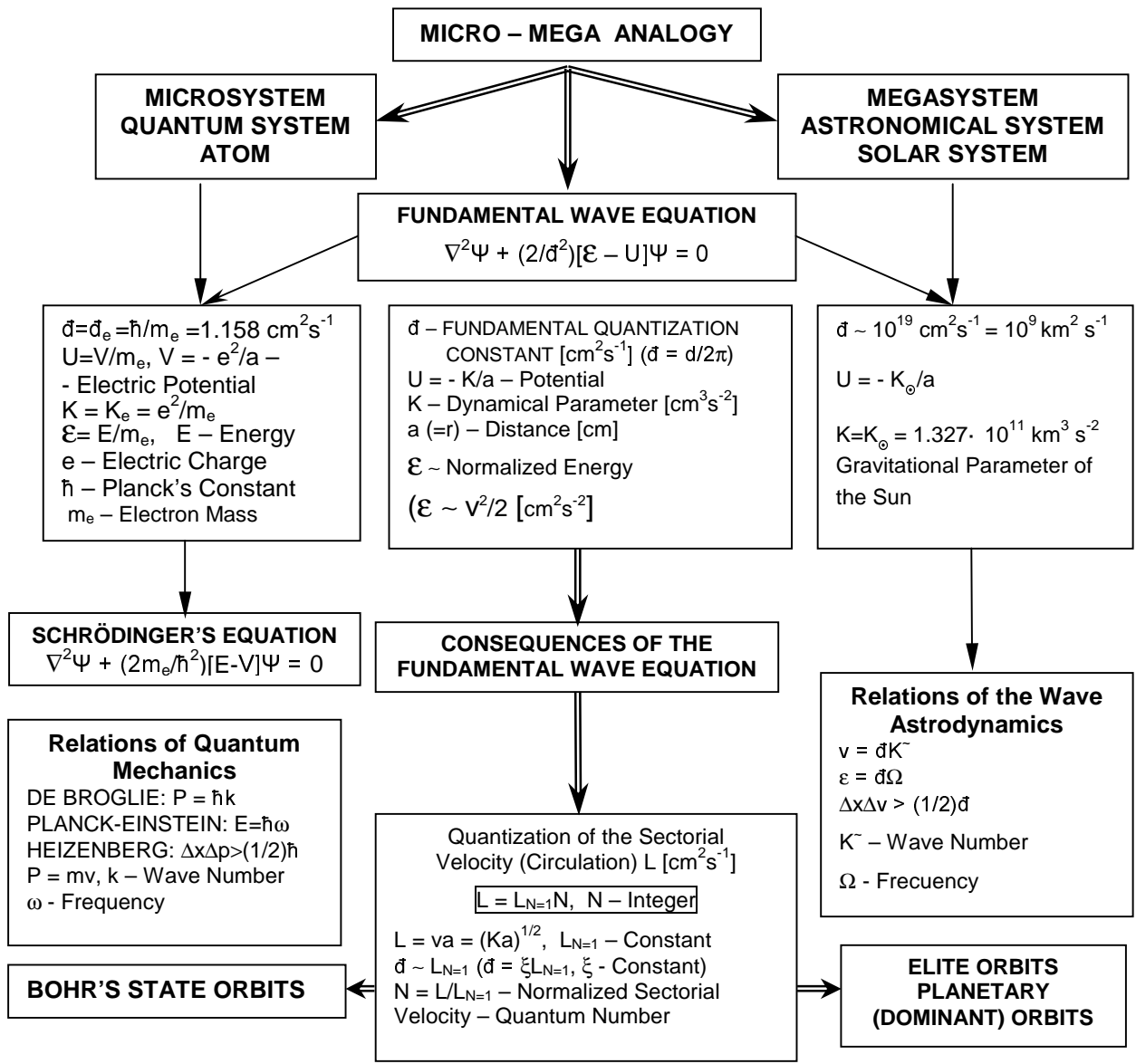